\documentclass[twocolumn,10pt]{IEEEtran}

\usepackage{amsfonts,amssymb,amsmath,calc,graphicx,subfigure}
\usepackage{psfrag,array,epsfig,pstricks,multicol,verbatim,moreverb,mathrsfs}
\usepackage{algorithm}
\usepackage{algorithmic}
\usepackage{color}

\newtheorem{theorem}{Theorem}

\newtheorem{corollary}{Corollary}
\newtheorem{definition}{Definition}
\newtheorem{assumption}{Assumption}
\newcommand{\qed}{\nobreak \ifvmode \relax \else
      \ifdim\lastskip<1.5em \hskip-\lastskip
      \hskip1.5em plus0em minus0.5em \fi \nobreak
      \vrule height0.5em width0.5em depth0.25em\fi}
\def\G{{\mathcal{G}}}
\def\P{{\mathcal{P}}}
\def\S{{\mathcal{S}}}

\def\R{{\mathcal{R}}}
\def\E{{\mathcal{E}}}
\def\V{{\mathcal{V}}}
\def\B{{\mathcal{B}}}

\def\M{{\mathcal{M}}}
\def\N{{\mathcal{N}}}
\def\T{{\mathcal{T}}}
\def\U{{\mathcal{U}}}
\def\X{{\mathcal{X}}}

\begin{document}
\title{Max-min Fairness in 802.11 Mesh Networks}
\author{Douglas J. Leith, Qizhi Cao, Vijay G. Subramanian\\Hamilton Institute, NUI Maynooth\thanks{This material is based upon works supported by the Science Foundation Ireland under Grant No. 07/IN.1/I901. }}
\maketitle

\begin{abstract}
In this paper we build upon the recent observation that the 802.11 rate region is log-convex and, for the first time, characterise max-min fair rate allocations for a large class of 802.11 wireless mesh networks.
\end{abstract}

\section{Introduction}

In this paper we build upon the recent proof in \cite{logconvexity10} that the 802.11 rate region is log-convex and, for the first time, characterise max-min fair rate allocations for a large class of 802.11 mesh networks.   By exploiting features of the 802.11e/n MAC, in particular TXOP packet bursting, we are able to use this characterisation to establish a straightforward, practically implementable approach for achieving max-min throughput fairness.   We demonstrate that this approach can be readily extended to encompass time-based fairness in multi-rate 802.11 mesh networks.

Fairness in 802.11 networks has been the subject of a considerable body of literature.  A large part of this literature is concerned with  \emph{unfairness} behaviour in 802.11 networks due to hidden terminals, exposed terminals, capture, upload/download unfairness \emph{etc.}, see for example \cite{macaw94,clifford07,liew08,Kochut,tcpfairness_2005} and references therein.   Proportional fairness over a single 802.11 hop is considered by \cite{siris06} , but this work makes the simplifying assumption that  every wireless station in a WLAN is always saturated, which cannot be expected to hold in general and is an unreasonable hypothesis for multi-hop networks.   An extensive literature relates to utility fairness in wired networks, but the CSMA/CA scheduling used in 802.11 differs fundamentally from wired networks due to carrier sense deferral of the contention window countdown and  the occurrence of colliding transmissions -- both of which act to couple together the scheduling of transmissions by stations in a WLAN and lead to the rate region being non-convex.    Utility fairness has been considered in random access wireless networks, but this work has been confined to the Aloha MAC, see \cite{tassiulas04,sasha06,calderbank06,wang04} and references therein.  The Aloha framework assumes that idle and transmission slots are of the same duration and so does not encompass standard 802.11 frame structure where (i) it is common for transmissions to be more than an order of magnitude longer than the idle slot duration in order to improve throughput efficiency and (ii) the mean transmission duration is not identical at all stations but instead depends on the packet size and PHY rate selected.  While it has been known for some time that Aloha networks have a log-convex rate region \cite{tassiulas04,wang04}, it has only recently been established that the 802.11 rate region is also log-convex \cite{logconvexity10}; it is this fundamental result that underpins the max-min utility fair analysis in the present paper.





\section{Network Model}

\subsection{Network Architecture}
\begin{figure}[bth]
\centering
\includegraphics[width=3.5in]{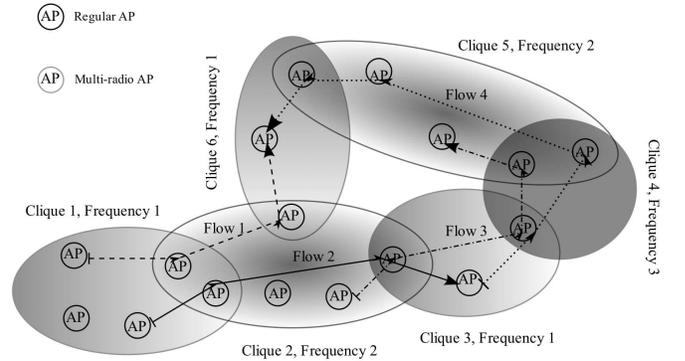}
\caption{Illustrating class of mesh networks considered.}\label{fig:meshfig}
\end{figure}

We consider a mesh network formed from a set of inter-connected WLANs and assume that the WLANs are non-interfering \emph{i.e.} that they either transmit on orthogonal channels or are physically separated so that transmissions on the same channel do not interfere.   Traffic is routed between WLANs via mesh points equipped with multiple radios.   Communication between mesh points is peer-to-peer so that sending a packet from WLAN $i$ to WLAN $j$ involves a single transmission (rather than routing via a central access point).   We assume that stations within a WLAN are within sensing distance of one another \emph{i.e.} there are no hidden terminals; we comment later on incorporating hidden terminals.  Such a mesh network is illustrated, for example, in Figure~\ref{fig:meshfig}.  In this example the network is formed from five inter-connected WLANs such that three orthogonal channels are sufficient to achieve a non-interfering allocation. 

\subsection{Station throughput}

Consider one of the WLANs in the mesh network and let $n$ denote the number of stations in the WLAN.  Following \cite{David_TON_2007},  we divide time into MAC slots where each MAC slot may consist either of a PHY idle slot, a successful transmission or a colliding transmission (where more than one station attempts to transmit simultaneously).  Let $\tau_i$ denote the probability that station $i$ attempts a transmission in a slot.  The mean throughput of station $i$ is then (\emph{e.g.} see \cite{David_TON_2007})
\begin{align*}
s_i (\T) = \frac{\tau_i \prod_{k=1,k\ne i}^n (1-\tau_k)D_i}{\sigma P_{idle} + T_sP_{succ}+T_c(1-P_{idle}-P_{succ})} 
\end{align*}
where $P_{idle}=\prod_{k=1}^n (1-\tau_k)$ is the probability that a slot is a PHY idle slot, $P_{succ}=\sum_{i\in N}\tau_i \prod_{k=1,k\ne i}^n (1-\tau_k)$ is the probability that a slot is a successful transmission, $\T=[\tau_1\ ...\ \tau_n]^T$ is the vector of attempt probabilities, $D_i$ is the mean number of bits sent by station $i$ in a successful transmission, $\sigma$ is the PHY idle slot duration, $T_s$ is the mean duration of a successful transmission (including time to transmit each data frame, receive the MAC ACK and wait for DIFS) and $T_c$ the mean duration of a collision.     

\subsection{Incorporating TXOP}
Later, we will make use of the TXOP packet bursting in 802.11e/n to facilitate achieving max-min fairness.    With TXOP, the length of time during which a station can keep transmitting without releasing the channel once it wins a transmission opportunity is specified as a control parameter.  In order not to release the channel, a SIFS interval is inserted between each packet-ACK pair and a successful transmission round then consists of multiple packets and ACKs.  By adjusting the TXOP time the number of packets that may be transmitted by a station at each transmission opportunity can be controlled.   We can readily generalise the above throughput expression to support TXOP packet bursting as follows.  Firstly observe that when TXOP packet bursting is used colliding transmissions end after sending the first packet in a burst and so $T_c$ is unchanged.  However, the duration $T_s$ of a successful transmission now depends on the size of the TXOP packet burst.   To encompass situations where stations may transmit different sized bursts on winning a transmission opportunity we let $T_{s,i}$ denote the mean duration of a successful transmission by station $i$.  The throughput of station $i$ is then
\begin{align*}
s_i (\T) = \frac{\frac{\tau_i}{1-\tau_i}P_{idle}D_i}{\sigma P_{idle} + \sum_{i=1}^nT_{s,i}\frac{\tau_i}{1-\tau_i} P_{idle}+T_c(1-P_{idle}-P_{succ})} 
\end{align*}
It will prove useful to work in terms of the quantity $x_i=\tau_i/(1-\tau_i)$ rather than $\tau_i$.  With this transformation we have that $P_{idle}=1/\prod_{k=1}^n (1+x_k)$ and $P_{succ}=\sum_{i=1}^n x_i/\prod_{k=1}^n (1+x_k)$ and so
\begin{align}\label{eq:tput}
s_i(x,N)=\frac{N_ix_i}{X(x,N)}\frac{L_i}{T_{c}}
\end{align}
where $N_i=T_{s,i}/T_c$, $L_i=D_i/N_i$ and
\begin{align}\label{eq:Xdef}
X(x,N)=a+\sum_{k=1}^n (N_k-1)x_i + \prod_{k=1}^n (1+x_k)-1
\end{align}
with $a=\sigma/T_c$, $x=[x_1,...,x_n]^T$ and $N=[N_1,...,N_n]^T$.     We also have that the mean fraction of time spent by station $i$ on successful transmissions is
\begin{align}\label{eq:time}
t_i(x,N)=\frac{N_ix_i}{X(x,N)}
\end{align}
which is simply a rescaling of the station throughput expression (\ref{eq:tput}).

In the foregoing we have implicitly assumed that packet losses only occur due to colliding transmissions \emph{i.e.}
\begin{assumption}\label{assump-1}
Packet losses from sources other than collisions can be neglected.
\end{assumption}
We discuss relaxing this assumption and including channel noise losses in Section \ref{sec:assumptions} below.  In addition, we will generally make the following assumption,
\begin{assumption}\label{assump0}
Frame transmissions are of duration $T_c$.  
\end{assumption}
A TXOP burst therefore consists of a sequence of frame transmissions each of duration $T_c$.  This assumption yields the useful technical benefit that the  collision duration $T_c$ is invariant with the attempt rates $x_i$ used in a WLAN -- if stations used frames of different duration then the duration of a collision would depend on the specific set of stations involved in a collision and so on the attempt rates $x_i$.   More importantly, however, it is also a natural assumption in the context of 802.11e where TXOP bursts are specified in terms of their duration in seconds (which, in turn, is motivated by consideration of time-based fairness when stations use different PHY rates).   With this assumption, $N_i$ can be interpreted as the mean number of transmissions in a burst and $L_i$ as the mean size, in bits, of the payload of each frame.  
 


\subsection{Constraining burst size}
Before proceeding, it is important to note that it is necessary to suitably constrain the size $N_i$ of allowed TXOP packet bursts.   To see this, say we let $N_i=\lambda n_i$ with $\lambda>0$, $n_i>0$ and look at the behaviour as $\lambda\rightarrow \infty$.   It can be verified that $ds_i/d\lambda$ equals
\begin{align*}
\frac{n_ix_i}{X}\left(1-\frac{\lambda\sum_{j=1}^n n_jx_j}{\sum_{j=1}^n (\lambda n_j-1)x_j +a+ \prod_{j=1}^n (1+x_j)-1} \right)\frac{L_i}{T_{c}}
\end{align*}
which can be seen to be strictly positive.  That is, increasing $\lambda$ (and so burst size) $always$ increases throughput.  In the limit, 
$
s_i \rightarrow \frac{n_ix_i}{\sum_{j=1}^n n_jx_j }\frac{L_i}{T_{c}}
$
as $\lambda\rightarrow \infty$. Observe that the idle time and collision time terms (which remain of finite duration) are washed out in the denominator and so the efficiency of the network is maximised subject to the fixed per packet overhead embodied by $L_i/T_c$. In effect, this says that any point strictly in the interior of the simplex $\{ s: \sum_{i=1}^n\tfrac{s_i T_c}{L_i} \leq 1\}$ is achievable by an appropriate choice of $N_i$s.
This high efficiency comes at the price of unbounded delays and so is not of practical interest.   Instead, to maintain bounded delay it is necessary to constrain the burst size and we let $\bar{N}_i$ denote the maximum burst size admissible at station $i$.  


\subsection{Finite-load}
It is useful to distinguish between the attempt probability $\tau_i$ and the attempt probability design parameter $\bar{\tau}_i$.   $\bar{\tau}_i$ is the probability that station $i$ considers making a transmission in a slot, but a transmission will not actually take place unless at least one packet is available to send.   It is the attempt probability $\tau_i$ which is relevant for the foregoing throughput expressions.    

When $\tau_i=\bar{\tau}_i$ a station is said to be \emph{saturated} and sends a packet at every transmission opportunity, otherwise it is \emph{unsaturated}.  For unsaturated stations the attempt probability $\tau_i$  depends jointly on the offered load and $\bar{\tau}_i$.   We will assume that when a station is unsaturated the throughput is equal to the offered load \emph{i.e.} that stations have sufficient buffering that queue overflow losses can be neglected when a station is unsaturated\footnote{Conservation of packets  then means that the mean throughput must equal the mean arrival rate.   Observe also that, by Loynes theorem \cite{loynes}, for sufficiently large buffering we have the intuitive property that a station will be unsaturated whenever the mean packet inter-arrival time is less than the mean service time.  }.  We also assume that the corresponding attempt probability $\tau_i$ is the just value that makes throughput expression (\ref{eq:tput}) equal the offered load, \emph{i.e.}  \begin{assumption}\label{assump0a}
Let $\U$ denote the set of unsaturated stations in a WLAN and $\S$ the set of saturated stations, with $\U\cup\S=\{1,2,\dotsc,n\}$ and $\U\cap\S=\emptyset$.   Let  $y_i=\bar{\tau}_i/(1-\bar{\tau}_i)$.  The attempt rate at a saturated station $j\in\S$ is then $y_j$ and the set of admissible station attempt rates is $\X=[\underline{x}_1,y_1]\times...\times[\underline{x}_n,y_n]$ with $\underline{x}_i=y_i$ for $i\in\S$ and $0$ otherwise.  Let $s_i$ denote the offered load at unsaturated station $i\in\U$.  If a solution $(x,N)\in\X\times\N$, where $\N=[1,\bar{N}_1]\times...\times[1,\bar{N}_n]$, exists to the throughput balance equations
\begin{align}
s_i=\frac{N_ix_i}{X(x,N)}\frac{L_i}{T_{c}}\ \forall\ i\in\U\label{eq:one}
\end{align}
then the offered load of $s_i$ can be serviced  by unsaturated station $i\in\U$ with 
the attempt rate $x_i$ and burst-size $N_i$ solving the balance equations.   
\end{assumption}

Note that for solutions to \eqref{eq:one} to exist it is necessary and sufficient that the set
\begin{align*}
C^\prime&=\Bigg\{(x,N)\in\X\times\N: \forall\ i\in\U\ s_i\leq\frac{N_ix_i}{X(x,N)}\frac{L_i}{T_{c}}
\Bigg\}
\end{align*}
be non-empty\footnote{
This is because solutions to \eqref{eq:one} are a superset of solutions to the optimization problem
$
\min_{(x,N)\in C^\prime}  \prod_{i=1}^N x_iN_i
$. 
From the proofs of Theorems \ref{thm:one} and \ref{lem:txop} (see later) this optimisation  
can be transformed into a convex problem such that all the constraints are satisfied with equality at the optimal solution(s), if the problem is feasible.
}


\subsection{Realisation in 802.11e/n}

Following the approach taken in Bianchi-like throughput models (\emph{e.g.} see \cite{David_TON_2007} and references therein), transmissions by an 802.11 station can be modelled as a renewal process, with renewals occuring after a successful transmission or discard.    The attempt probability can then be directly related to the 802.11 MAC parameters $CW_{min}$, $CW_{max}$, \emph{etc}.   For simplicity, we will hereafter assume that the attempt probability design parameter $\bar{\tau}_i$ can be freely selected.   However, this is not a fundamental requirement of our analysis and can be readily relaxed provided any constraints imposed on  $\bar{\tau}_i$ continue to yield a log-convex rate region; in particular, Theorem \ref{lem:txop} below carries over in the obvious way.   As an example of admissible constraints on $\bar{\tau}_i$, consider an 802.11 WLAN where we select $CW_{min}=CW_{max}=CW$, where $CW$ is an appropriate constant \emph{e.g.} 32.    Then $\bar{\tau}_i$ is constrained to take the single value $2/(CW-1)$ and the attempt probability $\tau_i$ can take values in $[0,2/(CW-1)]$ as the offered load on stations is varied.   By Theorem \ref{thm:one} in Section \ref{sec:logconvex}, the corresponding WLAN rate region is log-convex.   Indeed, we can constrain $\bar{\tau}_i$ to take any finite set of values (\emph{e.g.} corresponding to $CW$ taking powers of 2) since the resulting rate region is the intersection of the log-convex rate regions corresponding to each of the individual constraints on $\bar{\tau}_i$ and is therefore log-convex.

\subsection{Additional notation}
We represent the connectivity between WLANs via graph $\G$ with vertices $\V$ and edges $\E$. Each vertex in $\V$ corresponds to a WLAN and an edge exists between WLANs that can communicate.   Edges are labelled by the radio channel used.  Let $\N_i(c)$ denote the set of neighbours of WLAN $i$ on channel $c$ \emph{i.e.} a set of peering mesh points.   We will assume that each such set uses a channel with a unique label, but this is just a notational assumption and does not require that the physical channels are all different (in practice physical channels would be reused to exploit spatial multiplexing).  Since there are no hidden terminals,  peering mesh points form a clique \emph{i.e.} $\N_j(c)\cup\{j\}=\N_i(c)\cup\{i\}$ $\forall\ j\in \N_i(c)$ and we let $n(c)=|N_j(c)\cup\{j\}|$ denote the number of peers on channel $c$.  
Let $\P$ denote the set of network flows. Associated with each flow $p$ is a source client station and  route $r(p)$ (assumed loop-free) consisting of edges in $\G$ (\emph{i.e.} triples $(i,j,c)$, $j\in \N_i(c)$) traversed by the flow.   For notational simplicity we assume that flows do not start/finish at mesh points.  Let $\P_{i,j}(c)$ denote the set of flows $\{p: (i,j,c)\in r(p), p\in \P\}$ relayed from WLAN $i$ to WLAN $j$ on channel $c$, $\P_i(c)=\cup_{j\in \N_i(c)} \P_{i,j}(c)$ denote the set of all flows relayed by WLAN  $i$ and $\P(c)=\cup_{i\in\V} \P_i(c)$ denote the set of all flows relayed by peers on channel $c$.  

\section{Log-convexity of rate region}\label{sec:logconvex}

We begin by extending the log-convexity analysis in \cite{logconvexity10} to include TXOP packet bursting, and then use this to establish log-convexity of the mesh network rate region.   We present a new method of proof that makes use of theory of posynomials and geometric programming~\cite{boyd,BoydTutorial2007}.
%

\begin{definition} \emph{WLAN Rate Region}. The rate region of a WLAN is the set $\R$ of achievable throughput vectors $S(x,N)=[s_1\ ...\ s_n]^T$, with $i$'th element given by (\ref{eq:tput}), as the vector $x$ ranges over $\X=[0,\bar{x}_1]\times...\times[0,\bar{x}_n]$ and the vector $N$ ranges over $\N=[1,\bar{N}_1]\times...\times[1,\bar{N}_n]$.
\end{definition}
\begin{definition}\emph{Log-convexity}.  A set $C\in {\mathbb{R}}^n$ is convex if for any $s^1, s^2 \in C$ and $0\le \alpha \le 1$, there exists an $s^* \in C$ such that $s^*=\alpha s^1 + (1-\alpha) s^2$. A set $C$ is log-convex if the set $\log C := \{\log s: s\in C\}$ is convex.
\end{definition}
\begin{theorem}\label{thm:one}
The WLAN rate-region $\R$ is log-convex.
\end{theorem}
\begin{IEEEproof}
The throughput of station $i$ is given by
\begin{align*}
s_i = \frac{N_i x_i}{X(x,N)}
\end{align*}
where 
\begin{align*}
X(x,N) & =a+\sum_{j\in \M} N_j x_j + \prod_{j\in \M} (1+x_j) -1 -\sum_{j\in \M} x_j \\
& = a + \sum_{j\in \M} N_j x_j + \sum_{k=2}^n \sum_{A\subseteq \M, |A|=k} \prod_{j\in A} x_j. 
\end{align*}
and $\M=\{1,2,..,n\}$ denotes the set of stations in the WLAN.  Making a change of variables to $y_j=\log(x_j)$ and $\eta_j=\log(N_j)$, we have
\begin{align*}
\log(s_i) & = y_i+\eta_i - \log\bigg( a + \sum_{j\in \M} \exp(\eta_j+y_j) \\
& \qquad \qquad \qquad \qquad + \sum_{k=2}^n \sum_{A\subseteq \M, |A|=k} \exp\Big(\sum_{j\in A} y_j\Big)\bigg).
\end{align*}
with $y_j\in Y_j:=(-\infty, \log(\bar{x}_j)]$, $\eta_j\in E_j:=(0,\log(\bar{N}_j)]$.
Note that the right-hand-side is a concave function of $(y,\eta)$ since it is the transformed version of the reciprocal of a posynomial \cite{boyd}. Then the definition of a concave function implies that
\begin{align*}
C_i& =\bigg\{(\mu,y,\eta)\in {\mathbb{R}}^n \times \prod_{j\in \M} Y_j \times \prod_{j\in \M} E_j: \\
&  \qquad\;  \mu_i \leq y_i+\eta_i - \log\bigg( a + \sum_{j\in \M} \exp(\eta_j+y_j) \\
& \qquad \qquad \qquad \qquad \qquad \;\; + \sum_{k=2}^n \sum_{A\subseteq \M, |A|=k} \exp\Big(\sum_{j\in A} y_j\Big)\bigg) \bigg\}
\end{align*}
is a convex set. Therefore, $C=\cap_{i\in \M} C_i$ is also a convex set. The log rate-region is then the image of $C$ under the (linear) projection map that takes $(\mu,y,\eta)$ to $\mu$. Thus, the log rate-region is convex.
\end{IEEEproof}

We also have the following corollary that will prove useful later.  Let $\R'(\bar{p})$ denote the set of achievable throughput vectors $S(x,N)=[s_1\ ...\ s_n]^T$ as the vector $x$ ranges over $\X\cap\{x:\prod_{i=1}^n(1+x_i)\le \bar{p}\}$ and the vector $N$ ranges over $\N$.
\begin{corollary}\label{cor:logconvex} The constrained WLAN rate region $\R'(\bar{p})$ is log-convex for any $\bar{p}\ge1$.
\end{corollary}
\begin{IEEEproof}
We require $\bar{p}\ge1$ for $\R'$ to be non-empty. Now using the same transformation as in the proof of Theorem~\ref{thm:one}, the constraint that $\prod_{i=1}^n(1+x_i)\le \bar{p}$ translates to restricting attention to the following set
\begin{align*}
D& =\bigg\{(\mu,y,\eta)\in {\mathbb{R}}^n\times \prod_{j\in \M} Y_j \times \prod_{j\in \M} E_j: \\
& \qquad \qquad \qquad \qquad \qquad \sum_{j\in \M} \log(1+\exp(y_j)) \leq \log(\bar{p}) \bigg\}
\end{align*}
which is a convex set as a consequence of $\log(1+\exp(\cdot))$ being a convex function. The log rate-region is then $C\cap D$ which is convex, thus establishing the corollary.
\end{IEEEproof}
We note that the proof above can be readily extended to show that other constraints on $\tau$ (or $x$) and $N$ vectors also yield a convex set under our chosen transformation\footnote{For example, consider a constraint of the form $\sum_{i=1}^n x_i^2 \leq 1$.  Since the left-hand-side becomes $\log\left(\sum_{i=1}^n \exp(2 y_i)\right) \leq 0$, log-convexity continues to hold.  Similarly, the constraint $\sum_{i=1}^n \tau_i^{2} \leq 1$ can be transformed to $\tfrac{\tau_i}{1-\tau_i} \leq x_i$ for all $i$ and $\sum_{i=1}^n \tau_i^2 \leq 1$ with $x_i$ replacing $\tfrac{\tau_i}{1-\tau_i}$ in all the throughput formulae.  Since the first set of constraints can be transformed to $\tau_i+\tfrac{\tau_i}{x_i}\leq 1$ for all $i$, the constraints are now posynomial constraints in $(\tau,x)$ and log-convexity continues to hold.  
}.  Since the station transmission time (\ref{eq:time}) is simply a rescaling of the station throughput expression (\ref{eq:tput}) we also have the following result.
\begin{corollary}
The sets of feasible transmission times corresponding to rate regions $\R$ and $\R'$ are also log-convex.
\end{corollary}

A mesh network carries flows which traverse the component WLANs.   Let $\R(\G)$ denote the network rate region \emph{i.e.} the set of feasible flow throughputs.   
Since the throughput of unsaturated stations equals their offered load (see Assumption \ref{assump0a} and related discussion regarding buffering requirements), the network rate region is obtained by the appropriate intersection of the individual WLAN rate regions.  It follows immediately from the log-convexity of the component WLAN rate regions that the mesh network rate region is log-convex, \emph{i.e.} we have the following corollary.
\begin{corollary}\label{cor:logconvex_network} The mesh network rate-region $\R(\G)$ is log-convex.
\end{corollary}
\begin{IEEEproof}
We will once again use the property that convexity is preserved when taking intersections. Thus, it suffices to outline the key steps of the proof. Consider mesh point $i$ on channel $c$. Let ${\mathcal{P}}_i$ be the set of flows relayed by this mesh point. Using the transformation from the proof of Theorem~\ref{thm:one} the additional constraint that we need to satisfy is the flow-balance constraint at every mesh point \emph{i.e.}
\begin{align*}
& \log\bigg(\sum_{p\in {\mathcal{P}}_i}\exp(\tilde{s}(p))\bigg) \\ & \; \leq y_i+\eta_i - \log\bigg( a + \sum_{j\in \N_i(c)} \exp(\eta_j+y_j) \\
& \; \; \quad \qquad \qquad \qquad + \sum_{k=2}^{n_l(c)} \sum_{A\subseteq \N_i(c), |A|=k} \exp\Big(\sum_{j\in A} y_j\Big)\bigg) 
\end{align*}
where $\tilde{s}(p)=\log(s(p))$, which is again a convex constraint. Such constraints have to be satisfied for all the mesh points, and hence we get log-convexity for the entire rate-region.
\end{IEEEproof}
%

\section{Max-min throughput fairness}\label{sec:tput}

In this section we establish our main result, characterising max-min fair throughput allocations in the class of 802.11 mesh networks considered.  

\subsection{Assumptions}
Before proceeding we make the following assumptions.  We will relax all of these assumptions later, but they are useful for gaining initial insight into the nature of the max-min fair throughput allocation.

\begin{assumption}\label{assump1}\emph{PHY rate}.  All stations in the WLAN on channel $c$ use the same PHY rate for transmissions.
\end{assumption}
It follows from Assumption \ref{assump0} that stations use the same frame size \emph{i.e.} $L_i=L$.
\begin{assumption}\label{assump1a}\emph{Maximum burst-size}.  A station can transmit a maximum of one frame per flow at each successful transmission.  It follows that $N_i\le \bar{N}_i=|\P_i|$, where $|\P_i|$ is the number of flows carried by station $i$, and we have an additional constraint for each flow, namely $s(p) \leq \tfrac{x_i}{X(x,N)}\frac{L_i}{T_{c}}$ for flows $p\in\P_i$ carried by station $i$
\end{assumption}

Note that the additional constraint introduced here can once again be transformed to a log-convex constraint and therefore Corollary~\ref{cor:logconvex_network} still holds and the network rate-region is still log-convex.
%
\begin{assumption}\label{assump2}\emph{Attempt probability}. All stations in the WLAN on channel $c$ use the same attempt probability design parameter $y(c)=\bar{\tau}(c)/(1-\bar{\tau}(c))$.
\end{assumption}
 
Recall that $\bar{\tau}\in[0,1)$ is the transmission attempt probability when a station is saturated (always has a packet to send), but the actual attempt probability will be lower when a station is unsaturated.  Note that $y(c)$ need not be the same for every WLAN, but stations within a WLAN are assumed to use the same value of attempt parameter.  

The channel idle probability $P_{idle}(c)$ in the WLAN on channel $c$ is $1/\prod_{k}(1+x_k)$.  
\begin{assumption}\label{assump2b}\emph{Idle probability}. $\prod_{k}(1+x_k)\le \bar{p}(c)$.
\end{assumption}
This assumption involves no loss of generality as by selecting $\bar{p}(c)$ sufficiently large we can always ensure that the constraint is inactive.  Nevertheless, including this assumption allows us to also consider smaller values of $\bar{p}(c)$ as we will see later.   By Corollary \ref{cor:logconvex}, the rate region is log-convex for any value of $\bar{p}\ge 1$.

\subsection{Water-filling \& Bottleneck links}

Assumptions \ref{assump1}-\ref{assump2b} do not change the log-convexity of the network rate region and so we immediately have that a unique max-min rate allocation exists.  The network rate region also has the free disposal property \cite{boudec07} (same as coordinate-convexity) since each co-ordinate of the throughput vector is lower bounded by 0 and any non-zero feasible vector can always be decreased -- by scaling the attempt rate vector $x$ -- while staying within the rate region.   By \cite[Theorem 3]{boudec07} the max-min solution can therefore be found by water-filling.  

Recall the water-filling algorithm in \cite{boudec07}:

%
\begin{algorithmic}[1]
\STATE Let $\P^0=\P$, $\R^0=\R(\G)$, $n=0$
\STATE \textbf{do}
\STATE\label{mainstep} Find $\max_{}\ T^n$  s.t. $S_p=T^n\ \forall p\in \P^n$, $S \in \R^n$
\STATE $\R^{n+1}=\{S\in\R^n :  S_p\ge T^n\ \forall p\in \P^n\}$
\STATE $\P^{n+1}=\{p\in \P^{n} : \forall S\in\R^{n+1},\ S_p>T^n\}$
\STATE $n=n+1$ 
\STATE \textbf{until} $P^n=\O$
\end{algorithmic}
\noindent where $\P$ is the set of network flows, $\R(\G)$ denotes the network rate region (\emph{i.e.} the set of feasible flow throughputs), $S$ denotes the vector of flow throughputs and $S_p$ is the throughput of flow $p$ (element $p$ of vector $S$).
%
On termination of this water-filling algorithm, the remaining point in $\R^n$ is the max-min fair allocation of flow throughputs.   

Step \ref{mainstep} is the key step in the algorithm.   It finds the maximum throughput $T^n$ that the flows in set $\P^n$ may collectively use while remaining within the network rate region.  The flows whose throughput cannot be increased above $T^n$ are then removed from set $\P^n$, and step \ref{mainstep} repeated.    We can express step  \ref{mainstep} more explicitly in our wireless mesh network context as:
\begin{align}
&\max_{x,T^n}\ T^n \label{eq:stepab} \\
s.t.\ & s(p)=T^n\ \forall p\in \P^n \label{eq:stepaa}\\
&s(p)\le\frac{x_{k}(c)}{X(c)}\frac{L}{T_c(c)},\ \forall\ p\in\P,\ (k,\bullet,c)\in r(p)\label{eq:step3a}\\
&\sum_{p:(k,\bullet,c)\in r(p)}s(p)=\frac{N_k x_{k}(c))}{X(c)}\frac{L}{T_c(c)}\ \label{eq:step3c}\\
&x\ge 0 \\
&\prod_{k}(1+x_k)\le \bar{p}(c)\label{eq:step3b}
\end{align}
Constraints (\ref{eq:step3a})-(\ref{eq:step3b}) ensure that the vector of flow rates lies within the network rate region.

For all flows there exists an iteration $n$ such that the flow is eventually removed from set $\P^n$ because its throughput cannot be increased above $T^n$.  When a flow is removed the constraint (\ref{eq:step3a}) is necessarily tight (\emph{i.e.} it cannot be loosened by any choice of $x$ while respecting the other constraints) for some WLAN $c$.   We say that flow $p$ is \emph{bottlenecked} at this WLAN.   Our interest in bottlenecks stems from the following property, which follows immediately from these observations,
\begin{theorem}A throughput allocation is max-min fair if and only if every flow has a bottleneck.\end{theorem}
Observe also that all of the flows bottlenecked at the same WLAN $c$ have the same throughout (owing to constraint (\ref{eq:stepaa})), and this is strictly greater than the throughput of the other flows which traverse this WLAN but are not bottlenecked there.  We have therefore established that the well-known bottleneck property of max-min throughput allocations in wired networks also carries over to 802.11 mesh networks.

\subsection{Main result}

Surprisingly, despite the complex nature of the mesh network rate region (where flow rates are strongly coupled at each WLAN), we can obtain an almost complete characterisation of the max-min allocation of station attempt probabilities and burst sizes within each WLAN.  This makes use of the characterisation of the max-min allocation in terms of waterfilling and bottlenecks.   

Recall that we say that a  flow is \emph{saturated} if it has a packet available to send at every transmission attempt by the station, and is otherwise \emph{unsaturated}.   

\begin{theorem}\label{lem:txop}Under Assumptions \ref{assump-1}-\ref{assump2b}, the max-min fair throughput allocation within each WLAN possesses the following properties:
\begin{enumerate}
\item The attempt rate design parameter $y(c)\ge\bar{x}(c)$  in each WLAN where $\bar{x}(c)$ is the attempt rate that maximises the throughput of saturated flows.
\item Flows bottlenecked at the WLAN send one frame at every successful transmission made by the station.  When $y(c)=\bar{x}(c)$, all bottlenecked flows are saturated.  When $y(c)>\bar{x}(c)$ they are unsaturated.
\item Non-bottlenecked flows are always unsaturated.   
\end{enumerate}
\end{theorem}
\begin{IEEEproof} See Appendix~\ref{sec:app1}.
\end{IEEEproof}


The importance of Theorem \ref{lem:txop} is that it goes a long way to telling us how we might realise a max-min fair allocation in wireless mesh networks.  Specifically, consider a mesh network where each WLAN is configured as follows:
\begin{enumerate}
\item Stations in a WLAN all use the same attempt rate parameter $y(c)$ (\emph{e.g.} in 802.11 terminology, all stations in a WLAN use the same value of $CW_{min}=CW_{max}$).
\item Stations use per flow queueing and at each transmission opportunity send one frame of data from each non-empty queue.  
\item Parameter $y(c)$ is selected to maximise the throughput of saturated flows in  WLAN $c$.
\end{enumerate}
The network then satisfies Assumptions \ref{assump1}-\ref{assump2b}.  Observe that the per flow queueing discipline trivially ensures that $y(c)=\bar{x}(c)$ (saturated flows will transmit a packet at every transmission opportunity).  By Theorem \ref{lem:txop} we then have an equivalence between bottlenecked flows and saturated flows.   This equivalence is of fundamental importance.   Specifically, suppose each flow uses ideal congestion control \emph{\emph{i.e.}} adjusts the flow rate to ensure that the flow is saturated at one or more WLANs without incurring queue overflow losses.   Then congestion control will ensure that every flow is bottlenecked and so, without further effort, by Theorem  \ref{lem:txop}  the network throughput allocation will be max-min fair.   That is, we have the following important corollary of Theorem \ref{lem:txop}.
\begin{corollary}\label{cor:main}
Suppose each flow uses ideal congestion control and each WLAN in a mesh network is configured as stated above.   Then the resulting flow throughput allocation is max-min throughput fair.
\end{corollary}

Of course, in practice we must work with real rather than ideal congestion control.  Nevertheless, under suitable continuity conditions, we can expect that any congestion control algorithm that approximates ideal behaviour sufficiently closely will, by  Corollary \ref{cor:main}, yield a throughout allocation that is close to max-min fair and this is indeed confirmed in simulations, see Section \ref{sec:sims}.    

The network configuration in Corollary \ref{cor:main} also requires that attempt probability parameter $y(c)$ is selected to maximise the throughput of saturated flows in a WLAN.   This is considered in detail in the next section.  However, we note briefly here that the reason for introducing Assumption \ref{assump2b} is that by appropriately selecting $\bar{p}(c)$  then it turns out that $y(c)$ can be found in a completely decentralised manner (\emph{i.e.} no message-passing or packet-sniffing) using an approach similar to the idle-sense strategy for maximising WLAN throughput studied in \cite{idlesense}.   Assumption  \ref{assump2b} could alternatively be replaced by another constraint that simplifies selection of $y(c)$ so long as we retain log-convexity of the rate region.   For example, as noted earlier we could simply impose the constraint that $y(c)=y$ for an appropriate fixed value $y$, in which case no adaptation is required (this corresponds to trivially selecting $CW_{min}=CW_{max}=CW$ where $CW$ is some fixed value), although this appealing simplicity comes at the cost of a reduction in network capacity.

\section{Maximising throughput}
\begin{figure}
\centering
\includegraphics[width=0.6\columnwidth]{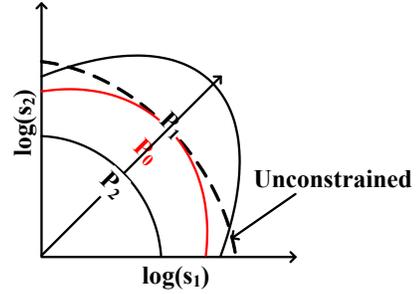}
\caption{Illustrating unconstrained rate region and rate region with $P_{idle}$ constraint.}\label{fig:pidle2}
\end{figure}

\subsection{Rate region boundary}
We begin by  studying the boundary of the rate region of WLAN $c$.   For this we will take a vector $y$, normalised such that $\sum_i y_i=1$, and set $x_i(c) = \lambda y_i/N_i$, $\lambda\ge 0$.  The vector of station throughputs is then $s=\lambda \frac{y}{X(c)} \frac{L}{T_{c}}$.  Since $\lambda$, $X(c)$, $L$ and $T_c(c)$ are all scalars it can be seen that varying $\lambda$ adjusts the position of the throughput vector on the ray in direction $y$ passing through the origin.   To determine the rate region boundary we need to find the values of $\lambda$ and $N_i$ that solve the optimisation 
\begin{align}
&\max_{\lambda, N_i} \frac{\lambda}{X} \nonumber \\
s.t.\ &\lambda\ge 0,\ 0\le N_i\le |\P_i|, i\in\{1,...n(c)\} \nonumber\\
&\prod_{i=1}^{n(c)}(1+\lambda y_i/N_i)\le \bar{p}(c)\label{eq:pidle}
\end{align}
Since the objective is strictly increasing in $N_i$ (as already noted) and constraint (\ref{eq:pidle}) becomes looser as $N_i$ increases, at the maximum  $N_i$ will lie on the constraint $|\P_i|$.   It can be verified by inspection of the second derivative that $\lambda/X$ is a concave function of $\lambda$ and so has a unique turning point.  To find the maximising value of $\lambda$, we observe that this will be determined either by constraint (\ref{eq:pidle}) becoming active or by the turning point of $\lambda/X$, whichever occurs first.   This is illustrated in Figure \ref{fig:pidle2} -- the dashed line marks the unconstrained rate region (\emph{i.e.} without constraint (\ref{eq:pidle})) and the solid curves mark the rate region boundary for different values of $\bar{p}$.  For a sufficiently small value of $\bar{p}$ it is the constraint (\ref{eq:pidle}) that determines the boundary of the rate region, see curve marked $P_2$ in the figure.

To determine the turning point of $ \lambda/X$, and so the unconstrained rate region boundary (marked by the dashed line in Figure \ref{fig:pidle2}), differentiating $\lambda/X$ with respect to $\lambda$ yields
$$
\frac{1}{X^2}\left(X - \lambda  (\sum_{i=1}^n y_i + \sum_{i=1}^n \frac{y_i}{N_i}\prod_{j\ne i} (1+\frac{\lambda y_i}{N_i})-\sum_{i=1}^n\frac{y_i}{N_i})\right)
$$
 and setting this derivative equal to zero we have that the $\lambda^*$ corresponding to the turning point solves
 $$
 \sum_{i=1}^n \frac{\lambda^* y_i}{N_i}\prod_{j\ne i} (1+\frac{\lambda^* y_i}{N_i})+1-a = \prod_{i=1}^n(1+\frac{\lambda^* y_i}{N_i})
$$
Substituting, we therefore have that the turning point (\emph{i.e.}, boundary of the rate-region) satisfies
\begin{align*}
\sum_{i=1}^n \frac{x_i^*}{1+x_i^*}\prod_{j=1}^n(1+x_j^*)+1-a=\prod_{j=1}^n(1+x_j^*)
\end{align*}
This can be rewritten as
\begin{align*}
\sum_{i=1}^n \tau^*_i + (1-a)P_{idle}^* = 1
\end{align*}
where $P_{idle}^*=\prod_{i=1}^n (1-\tau_i^*)$. Note that this is a generalization of the result
from \cite{MasseyMathys,Post} to the scenario with different slot lengths (\emph{i.e.}, $a<1$) and TxOP. 

Using the Arithmetic Mean-Geometric Mean inequality, we have
$$\sum_{i=1}^n\frac{1}{1+x_i} \geq n\sqrt[n]{\prod_{i=1}^n(\frac{1}{1+x_i})}$$
\emph{i.e.}
$$1-\frac{1}{\sqrt[n]{\prod_{i=1}^n (1+x_i)}} \geq \frac{1}{n}\sum_{i=1}^n \frac{x_i}{1+x_i}$$
After some algebra, it follows that selecting $\bar{p}(c) \le 1/(1+a-\sqrt{2a})$ ensures that constraint (\ref{eq:pidle}) is guaranteed to become tight either before or at the turning point of $\lambda/X$.  Note that $1+a-\sqrt{2a}\le 1$ when $a\le 1$ and this bound on $\bar{p}(c)$ is tight (with equality along the ray $\lambda\textbf{1}$, where $\textbf{1}$ denotes the all 1's vector, as $n\rightarrow \infty$).   This is illustrated by the middle curve marked $P_0$ in Figure \ref{fig:pidle2}, which touches the unconstrained rate region along the 45 degree ray.   With this choice of  $\bar{p}(c)$ constraint (\ref{eq:pidle}) is active at the solution to the above optimisation and so it is this constraint that determines the maximum value of $\lambda$, and thereby the maximum throughput of saturated flows.

\subsection{Decentralised optimisation}

Recall that our task is to select attempt rate parameter $y(c)$ to maximise the throughput of saturated flows.   Selecting $\bar{p}=1/(1+a-\sqrt{2a})$ so as to maximise the constrained rate region, it follows from the discussion in the preceding section that the throughput of saturated flows is maximised when $P_{idle}(c)= 1/\bar{p}(c)$.   That is, we need to select $y(c)$ such that $P_{idle}(c)= 1/\bar{p}(c)$.   This can be achieved in an entirely decentralised manner since (i) the idle probability $P_{idle}(c)$ can be directly observed by all stations in a WLAN (via carrier-sense, see for example \cite {clifford07}) and (ii) algorithms such as AIMD can be used to ensure stations converge to using the same parameter $y(c)$, see for example \cite{idlesense}.

\subsection{Degree of sub-optimality}
\begin{figure}
\centering
\includegraphics[width=0.9\columnwidth]{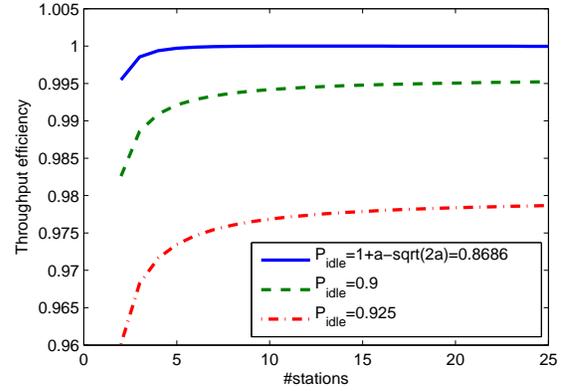}
\caption{Cost of operating on $P_{idle}=1+a-\sqrt{2a}$ constraint vs number of stations in WLAN. $a=1/100$, each station carries a single saturated flow.}\label{fig:pidle}
\end{figure}

Using any non-zero value of $1/\bar{p}$ necessarily comes at the cost of a reduction in throughput.   To see this note that when only a single station is active in a WLAN, and so no collisions are possible, then we ought to select the attempt probability  equal to 1 (\emph{i.e.} $y(c)\rightarrow \infty$) in order to maximise the throughout, in which case any value of $1/\bar{p}$ greater than zero must reduce throughput below its maximum value.  Nevertheless, the throughput loss is generally small.  For example,  Figure \ref{fig:pidle} illustrates the throughput cost of selecting $\bar{p}$ to ensure operation on the $P_{idle}(c)= 1/\bar{p}(c)$ constraint.   The figure plots the ratio of the throughput when  $P_{idle}(c)= 1/\bar{p}(c)=1+a-\sqrt{2a}$ to the maximum possible throughput when there is no $P_{idle}$ constraint.   It can be seen that the throughput efficiency is remarkably high, with a throughput reduction of less the 0.5\% (compared to the maximum possible throughput) even when only a single station is active.   This is similar to the observation made in \cite{idlesense}.   In return for this small cost we gain the advantage of a fully decentralised implementation with no message-passing.  The final choice of whether the additional network capacity to be gained by message-passing warrants the additional complexity lies with the network designer.

%

\section{Simulation results}\label{sec:sims}

\begin{figure}[bt]
\centering
 \subfigure[Topology]{\includegraphics[width=\columnwidth]{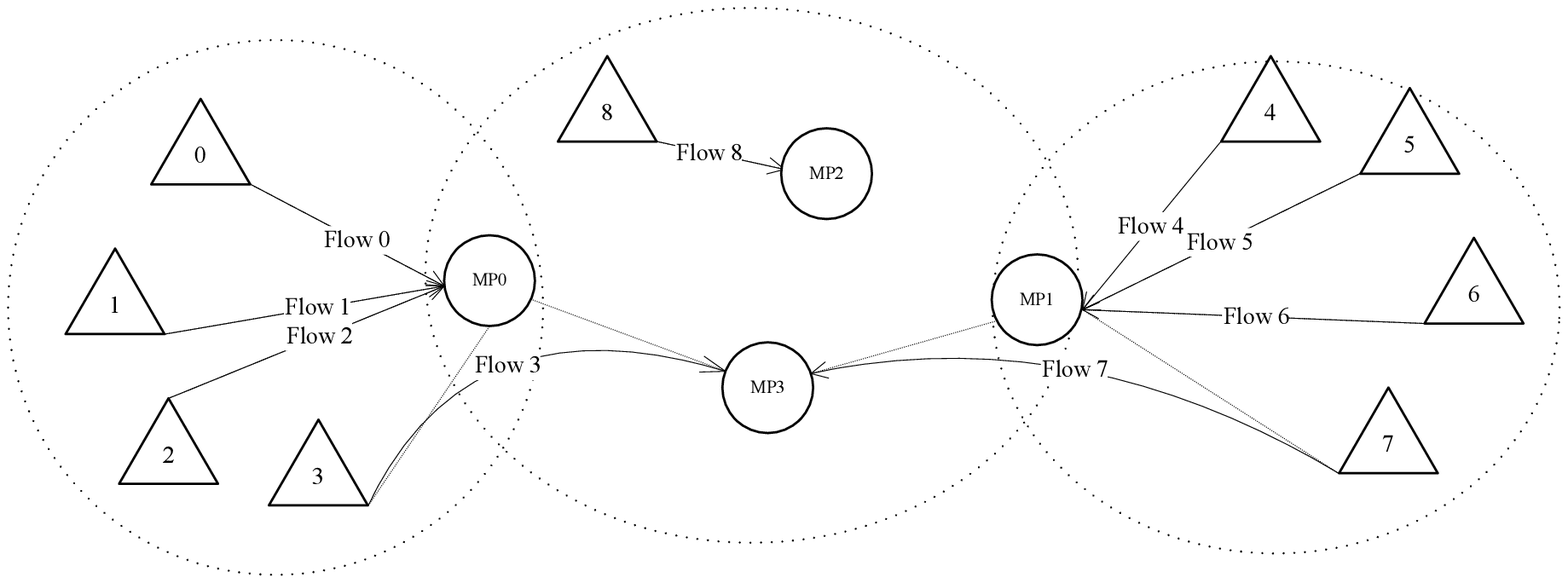}}
 \subfigure[Throughput Allocation]{\includegraphics[width=0.7\columnwidth]{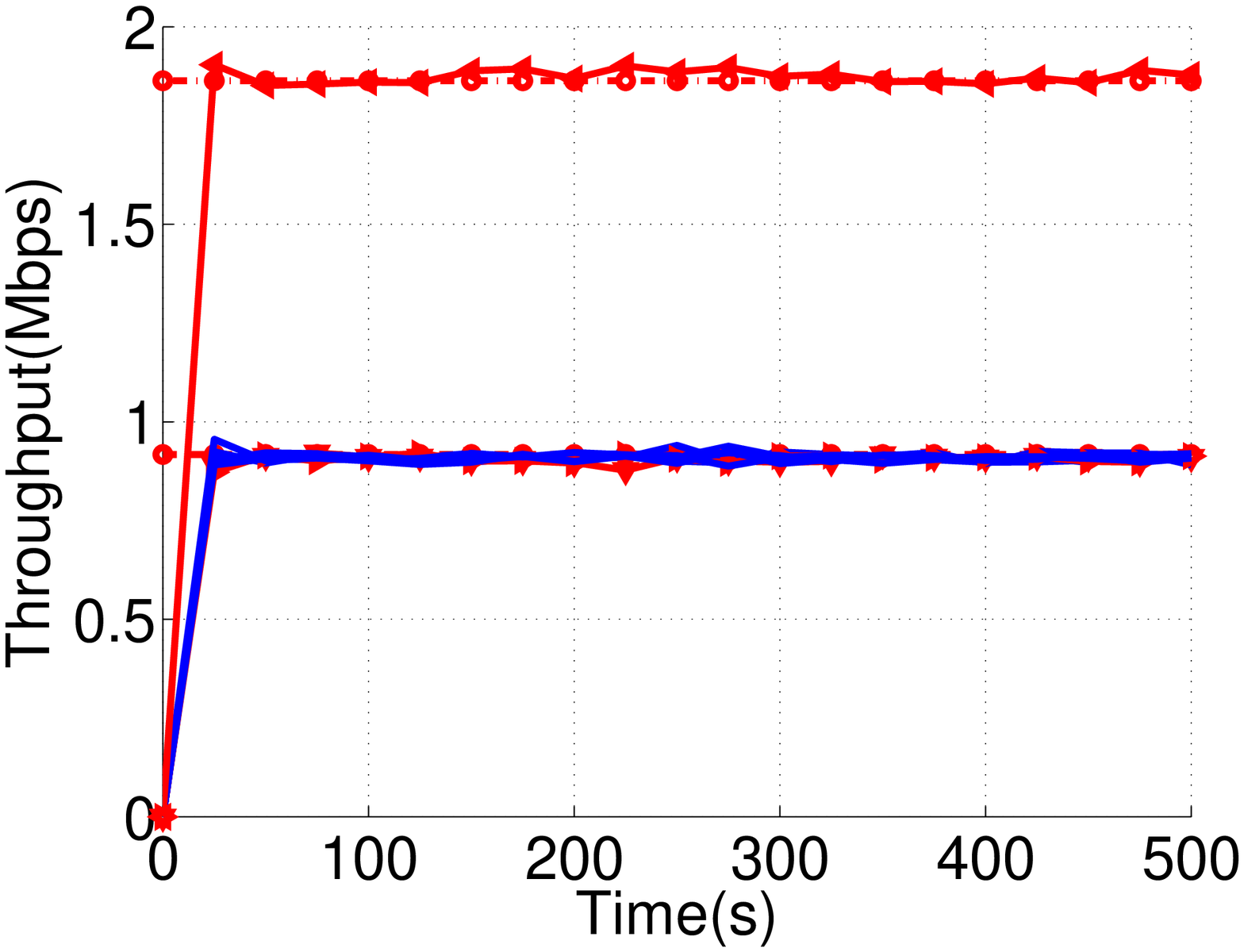}}
\caption{Example 1.  The measured simulation throughputs are compared against the theoretical max-min fair throughput values in the lower plot.   The lower valued lines are the throughputs for flows 0-7, while the upper valued lines are for flow 8.  The measurement points plotted are averages over 50s time windows.  The theory values for the bottlenecks are 0.92127Mbps and 1.8479Mbps respectively (indicated by the dashed red lines and also marked on y-axis by circles). It can be seen that the simulation values are in good agreement with theory. }\label{fig:ex1}
\end{figure}

We illustrate the foregoing analysis via $ns2$ packet-level simulations.  We begin by considering a mesh network with the topology shown in Figure \ref{fig:ex1}(a).   Mesh points (MP) are marked by circles and client stations by triangles.   Each WLAN operates on an orthogonal channel and MP0, MP1 are equipped with two radios to allow relaying of traffic between WLANs.   Flows 0-2 travel one hop to MP0, flow 3 travels two hops to MP3, flows 4-7 travel one hop to MP1, flow 7 travels two hops to MP3.   Flow 8 travels one hop from station 8 to MP2.  

In the simulations all flows are long-lived TCP traffic and so are bidirectional (\emph{i.e.} consisting of TCP data and TCP ACK packets).   Following \cite{tcpfairness_2005}, TCP ACKs are prioritised so that their loss rate is negligible (link asymmetry leading to excessive loss of TCP ACKs is well known to induce unfairness due to disruption of ACK clocking and repeated TCP timeouts).  The TCP ACK transmit time (including MAC ACK \emph{etc}) is lumped in with the TCP DATA transmit time to obtain the $T_s$ value for throughput formula (\ref{eq:tput}).  See \cite{wirelesscomm} for a more detailed discussion of the accuracy of this approximation, but we note here the good agreement in Figure \ref{fig:ex1}(b) between the theory values derived using this assumption and the simulation measured throughputs. 
  
\begin{figure}
\centering
 \subfigure[$CW_{min}$]{\includegraphics[width=0.7\columnwidth]{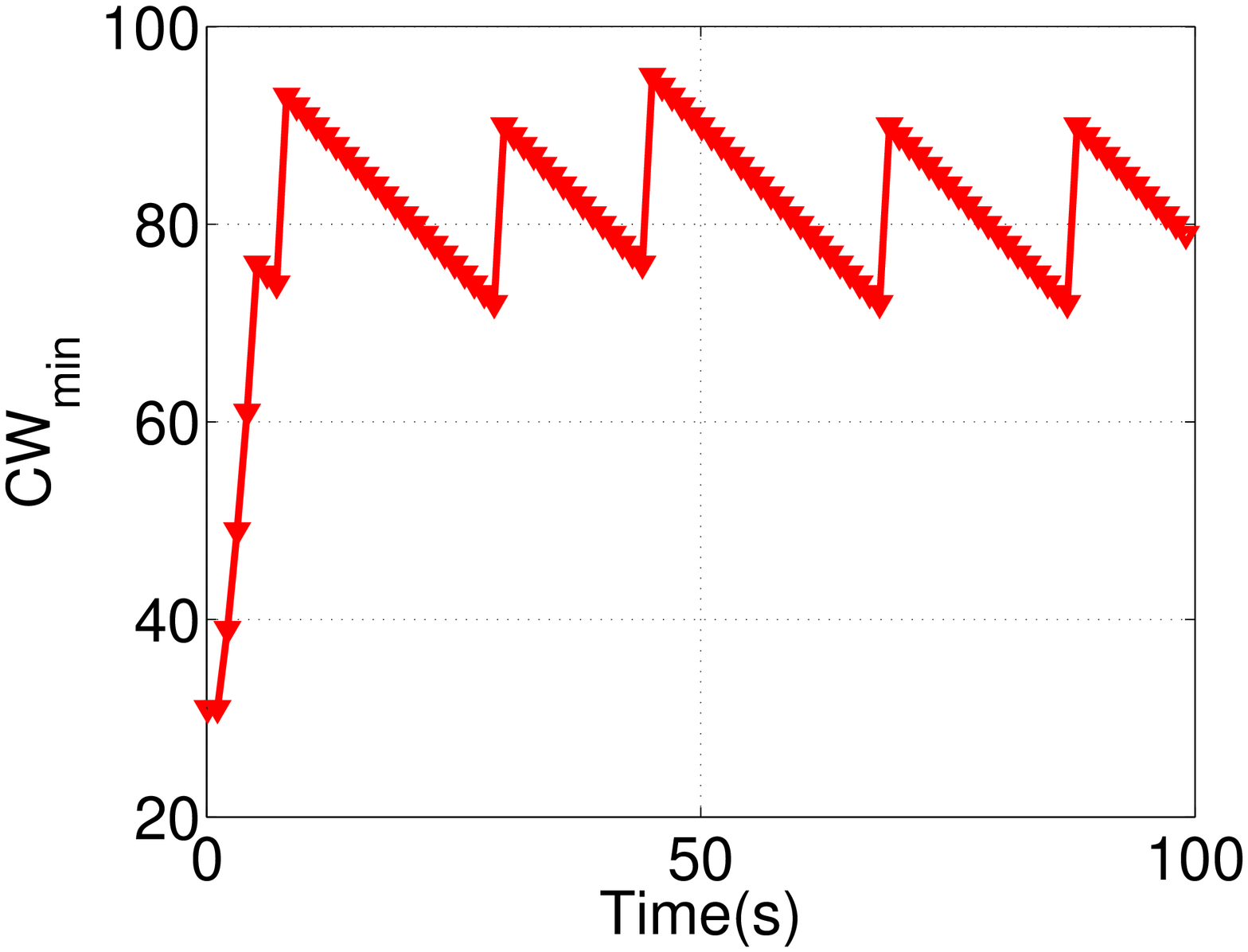}}
 \subfigure[$P_{idle}$]{\includegraphics[width=0.7\columnwidth]{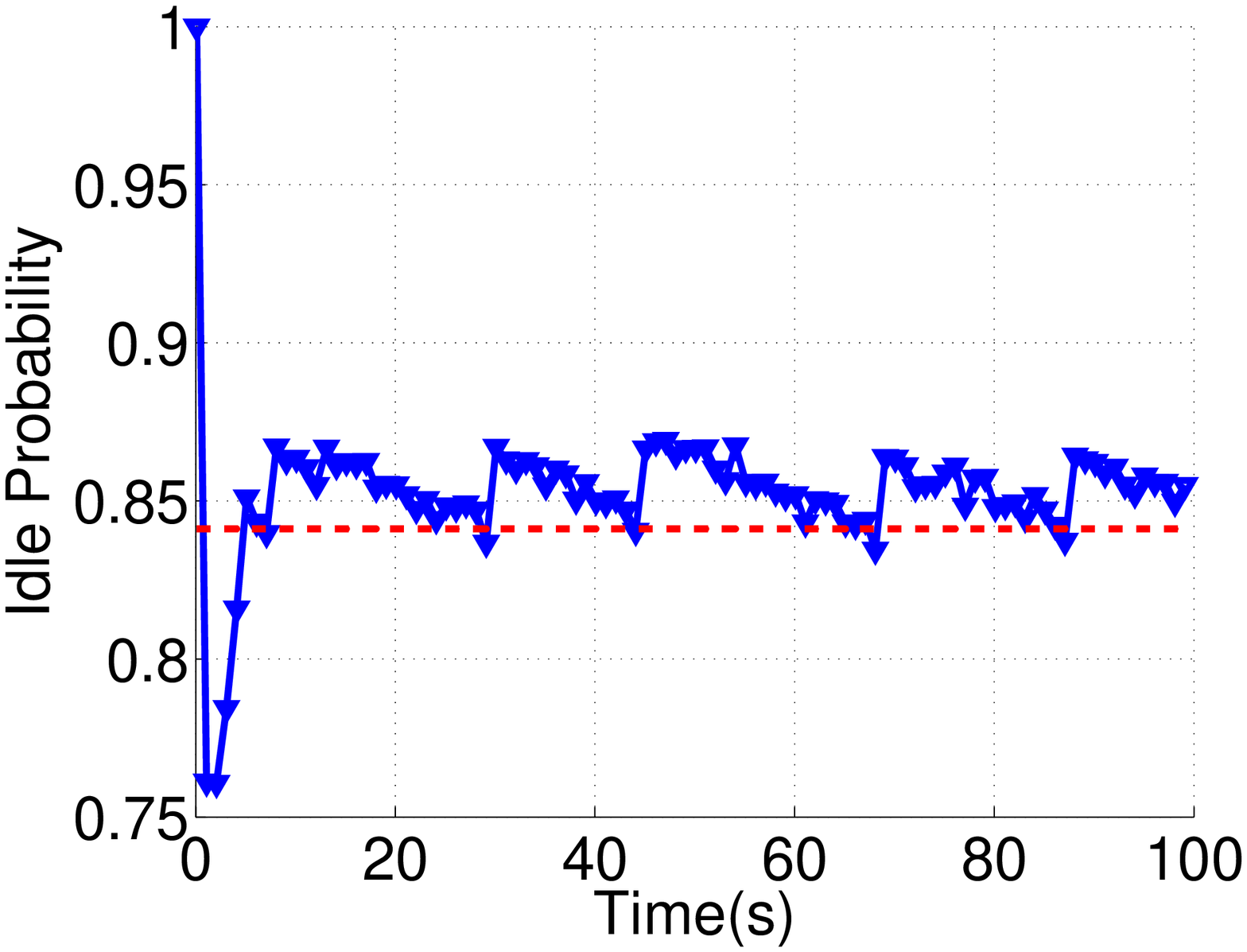}}
\caption{$CW_{min}$ and WLAN $P_{idle}$ time histories for the source station of flow 0 in the topology of Figure \ref{fig:ex1}(a).  These are representative of the time histories for other stations and illustrate the AIMD adjustment of $CW_{min}$.  The dashed line in the lower plot indicates the ideal $P_{idle}$ constraint value.}
\label{fig:ex1-cw}
\end{figure}

The stations in each WLAN measure the idle probability $P_{idle}$ using their carrier-sense functionality (\emph{e.g.} see \cite {clifford07}) and run a local AIMD algorithm to adjust their $CW_{min}$ to satisfy the constraint $P_{idle}\ge1/\bar{p}=1+a-\sqrt{2a}$, see Algorithm \ref{algo:aimd} for details.   Due to the use of the AIMD algorithm the station $CW_{min}$'s vary over time in a sawtooth pattern and do not settle on a constant value, see Figure \ref{fig:ex1-cw}(a).  Moreover, $CW_{min}$ is restricted to take integer values thereby introducing further granularity.    By adjusting the AIMD $\beta$ parameter the amplitude of the $CW_{min}$ sawtooth can be changed.   Decreasing $\beta$ reduces the size of the $CW_{min}$ fluctuations, but this comes at the cost of slower convergence to steady-state operation, \emph{e.g.} see \cite{ToN_BoB} for a detailed analysis of AIMD dynamics.   We choose $\beta=0.25$ as a compromise between fast convergence and reasonably small fluctuations in $CW_{min}$.  Due to these implementation issues, as can be seen from Figure \ref{fig:ex1-cw}(b), the WLANs do not operate exactly on the $P_{idle}=1/\bar{p}$ constraint as assumed in the calculation of the theoretical throughput values shown in Figure \ref{fig:ex1}(b).    Nevertheless, as can be seen from Figure \ref{fig:pidle} the throughput efficiency is relatively insensitive to $P_{idle}$ fluctuations around the optimum value and this is reflected in the good agreement between the theory and simulation throughputs in Figure \ref{fig:ex1}(b) .

\begin{figure}
\centering
\includegraphics[width=0.7\columnwidth]{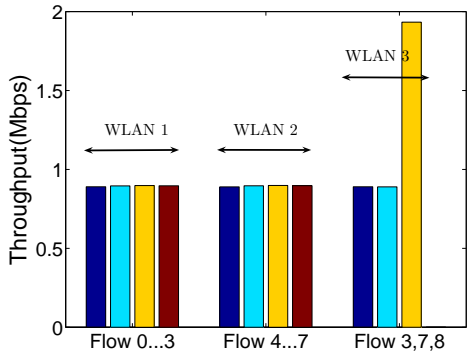}
\caption{Histogram of flow throughputs in each WLAN in topology Figure \ref{fig:ex1}(a).  WLAN 1 refers to the left-hand WLAN, WLAN 2 to the right-hand WLAN and WLAN 3 to the centre WLAN in Figure \ref{fig:ex1}(a).}
\label{fig:ex1-hist}
\end{figure}

Other simulation parameters used are detailed in Table \ref{simu_para}.   Figure \ref{fig:ex1}(b) compares the theoretical max-min fair throughout allocation with the measured simulation throughputs.  It can be seen that they agree remarkably well.   We can investigate the structure of the throughput allocation in the simulations in more detail.   By inspection of the topology in Figure \ref{fig:ex1}(a) we expect that the max-min throughput allocation has flows 0-3 bottlenecked at the left-hand WLAN, flows 4-7 at the right-hand WLAN and flow 8 at the centre WLAN.   Figure \ref{fig:ex1-hist} plots the flow throughputs in each WLAN, from which it can be seen that flows 0-3 are indeed the maximal throughput flows in the left-hand WLAN and similarly for flows 4-7 and flow 8 in the right-hand and centre WLANs respectively.   By inspection of the station queue occupancies (not plotted here), we can also confirm that flows 0-3 are saturated in the left-hand WLAN, and similarly for flows 4-7 and flow 8 at their respective bottlenecks, in accordance with Theorem \ref{lem:txop}.

\begin{figure}
\centering
 \subfigure[Topology]{\includegraphics[width=\columnwidth]{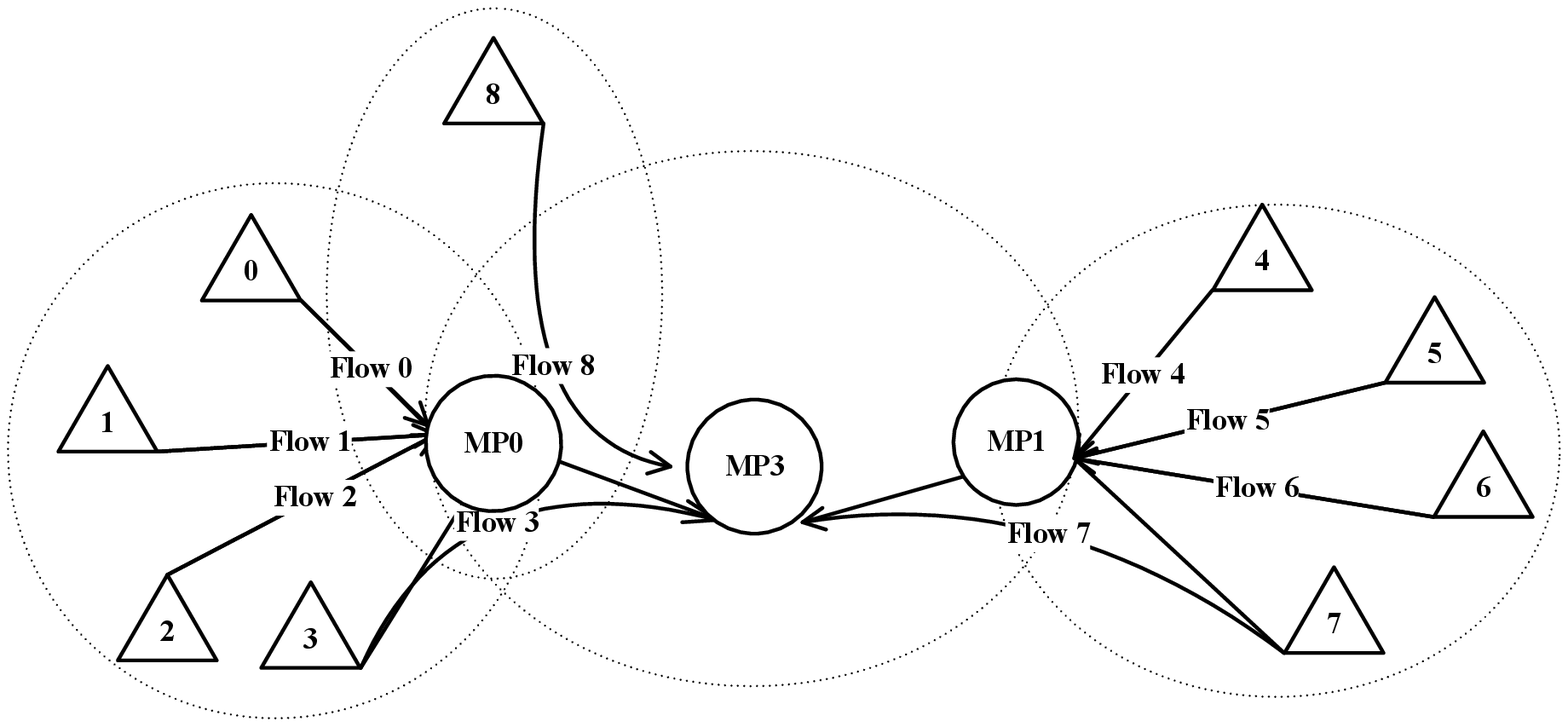}\label{fig_less_coll}}
 \subfigure[Throughput Allocation]{\includegraphics[width=0.7\columnwidth]{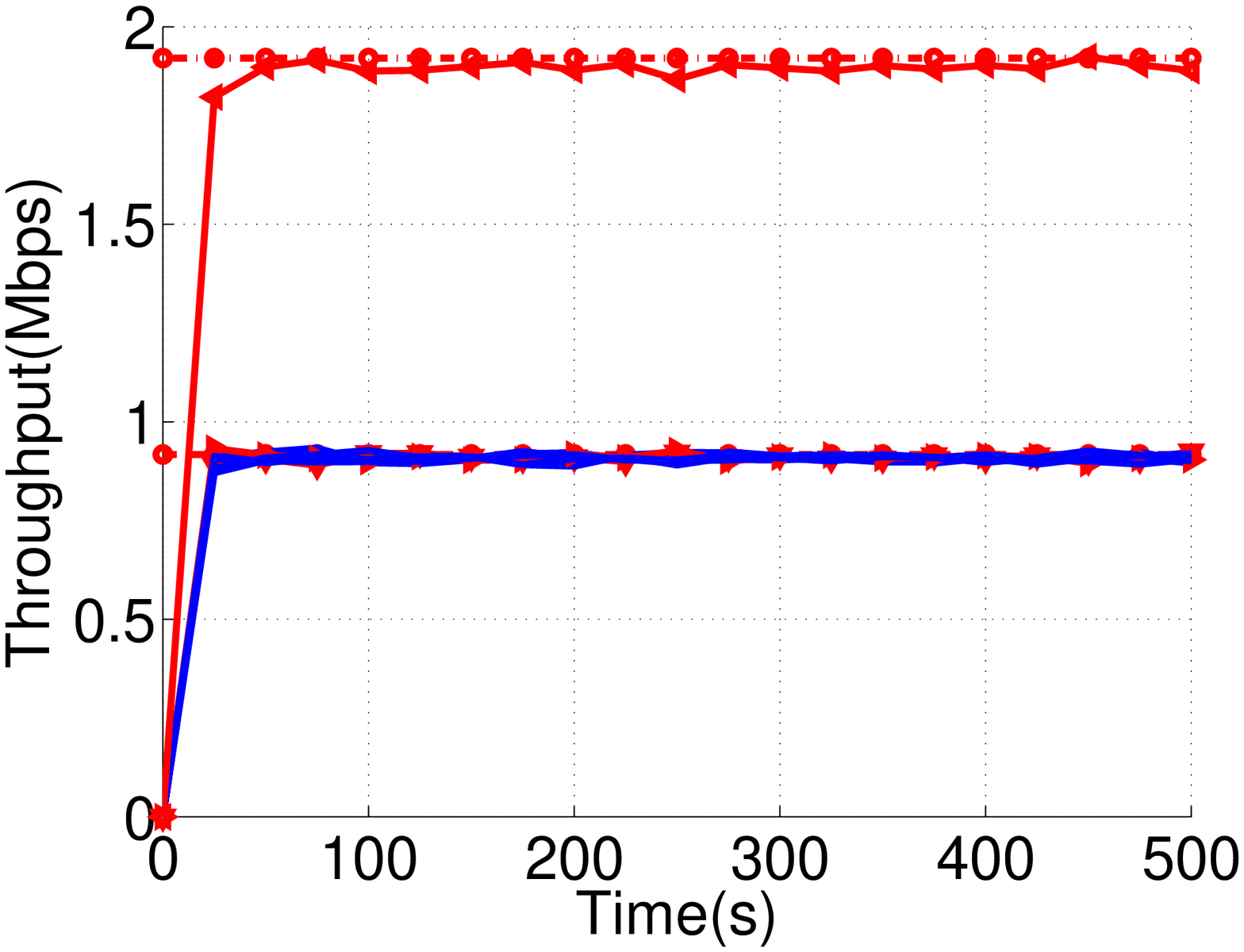}}
\caption{Example 2.  As before, the measured simulation throughputs and theoretical max-min fair throughput values (indicated by dashed red lines and also marked on y-axis by circles) are compared in the lower plot.}\label{fig:ex2}
\end{figure}

Figure \ref{fig:ex2} shows simulations results for a second topology.   An additional WLAN has been added containing station 8 and MP0 now carries two flows, namely flow 3 and flow 8.   Flow 8 is bottlenecked at the link between MP0 and MP3 while flow 3 is not, and simulations confirm that flow 8 is saturated at MP3 while flow 3 is not as per Theorem \ref{lem:txop}.   Also note that in this modified topology the one-hop flow 8 is allocated a slightly higher throughput than in Fig~\ref{fig:ex1} because there are now fewer collisions in the centre channel which is the bottleneck for this flow  -- MP0 and MP1 are transmitting data packets and MP3 transmitting TCP ACK packets, while in Fig~\ref{fig:ex1} we additionally have traffic between station 8 and MP2 in this channel.   Once again, observe that the simulation measurements agree extremely well with the theoretical max-min throughput allocation.
 

\begin{algorithm}[tb]
\begin{algorithmic}[1]
\FOR {Every T seconds}
\STATE Check the measured idle probability $P_{idle}$
\IF { $P_{idle} > 1/\bar{p}$ }
\STATE $CW_{min} \leftarrow CW_{min} + \alpha$
\ELSE
\STATE $CW_{min}\leftarrow CW_{min}\times(1-\beta)$
\ENDIF
\ENDFOR
\end{algorithmic}\label{algo:aimd}
\caption{AIMD algorithm used at each station to adjust its $CW_{min}$ value.}
\end{algorithm}

\begin{table}
\centering
\begin{tabular}{|c|c|}
\hline
PHY rate (Mbps)&11\\\hline
NIC Buffer (Packets) &50\\\hline
Packet Length (Bytes)&1000\\\hline
$\alpha$&4\\\hline
$\beta$&0.25\\\hline
T(s)&1\\\hline
$1/\bar{p}$&0.8412\\\hline
\end{tabular}
\caption{Simulation Parameters}
\label{simu_para}
\end{table}


\section{Time-based max-min fairness}\label{sec:timebased}

We can readily extend the foregoing analysis to encompass weighted max-min fairness, \emph{i.e.} where rather than max-min fairness of the flow throughputs $s(p)$, $p\in\P$ we require max-min fairness of the weighted flow throughputs $s(p)/w(p)$, $p\in\P$ for specified weights $w(p)>0$.   This is of particular interest when we relax Assumption \ref{assump1} that stations within a WLAN use the same PHY rate.   When flows can use different PHY rates, max-min throughput fairness leads to flows with a low PHY rate grabbing bandwidth from higher PHY rate flows, potentially leading to a large reduction in network capacity.    Time-based fairness is therefore typically of greater interest than throughout fairness in multi-rate networks, \emph{e.g.} see \cite{idlesense,tan04,liew05} and references therein.   Let $R(p)$ denote the PHY rate used by flow $p$, which for simplicity we assume is the same at every hop along the flow route $r(p)$.   The airtime used by flow $p$ is then given by $t(p)=s(p)/R(p)$ and so time-based fairness corresponds to weighted max-min fairness with weights $w(p)=R(p)$.  

Since the airtime is  just a rescaling of the throughput it follows that the  feasible set of times is log-convex and a unique max-min time allocation exists.  Retaining Assumptions \ref{assump1a}-\ref{assump2b} (for the moment), step \ref{mainstep} of the water-filling algorithm becomes

 \begin{align*}
&\max_{x,T^n}\ T^n \\
s.t.\ & t(p)=T^n\ \forall\ p\in \P^n \\
&t(p)\le\frac{x_{k}(c)}{X(c)},\ \forall\ p\in\P,\ (k,\bullet,c)\in r(p)\\
&\sum_{p:(k,\bullet,c)\in r(p)}t(p)=\frac{N_k x_{k}(c))}{X(c)}\ \\
&x\ge 0 \\
&\prod_{k}(1+x_k)\le \bar{p}(c)
\end{align*}
An identical argument to that used in the proof of Theorem \ref{lem:txop} can be applied (since $L/T_c(c)$ is just a constant scaling in the expressions used in the proof) to obtain 
\begin{theorem}\label{cor:time}Under Assumptions \ref{assump-1}-\ref{assump0a},\ref{assump1a}-\ref{assump2b}, the max-min fair time allocation within each WLAN possesses the following properties:
\begin{enumerate}
\item The attempt rate design parameter $y(c)\ge\bar{x}(c)$  in each WLAN where $\bar{x}(c)$ is the attempt rate that maximises the throughput of saturated flows.
\item Flows bottlenecked at the WLAN send one frame at every successful transmission made by the station.  When $y(c)=\bar{x}(c)$, all bottlenecked flows are saturated.  When $y(c)>\bar{x}(c)$ they are unsaturated.
\item Non-bottlenecked flows are always unsaturated.   
\end{enumerate}
\end{theorem}

It can be seen that the properties of the max-min time allocation are \emph{identical} to those of the max-min fair throughput allocation with a single PHY rate and so the same network configuration (together with ideal congestion control) can be used to realise the max-min time allocation \emph{i.e.}
\begin{enumerate}
\item Stations in a WLAN all use the same attempt rate parameter $y(c)$.
\item Stations use per flow queueing and at each transmission opportunity send one frame from the head of each non-empty queue (recall that by Assumption \ref{assump0} that all frames are of equal duration, regardless of the PHY rate used).  
\item Parameter $y(c)$ is selected to maximise the throughput of saturated flows in  WLAN $c$.
\end{enumerate}


\subsection{Simulation results}

\begin{figure}[bt]
\centering
\subfigure[Throughput Allocation]{\includegraphics[width=0.7\columnwidth]{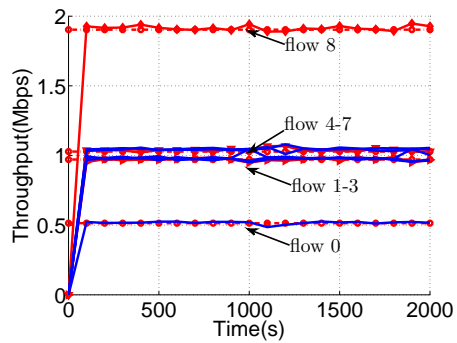}}
\subfigure[Time Allocation]{\includegraphics[width=0.7\columnwidth]{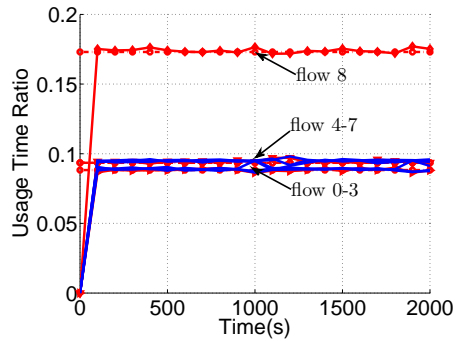}}
\caption{Multi-rate variant of Example 1.  Flow 0 uses PHY rate of 5.5Mbps, other flows a rate of 11Mpbs.  Plots compare simulation measurements and theoretical values (indicated by dashed red lines and also marked on y-axis by circles) of a max-min fair time allocation.  The measurement points plotted are averages over 50s time windows. }\label{fig:ex3}
\end{figure}

We revisit the previous simulation example in Figure \ref{fig:ex1}, but now extend consideration to a multi-rate situation where flow 0 in the left-hand WLAN uses a PHY rate of 5.5 Mbps while all other flows in the mesh network use a PHY rate of 11 Mbps.    Figure \ref{fig:ex3} compares simulation measurements with theoretical values for a max-min fair time allocation.   It can be seen from Figure \ref{fig:ex3}(a) that  flow 0 (the flow with lower PHY rate) is now allocated a lower throughput than the other flows in the left-hand WLAN.   This ensures that all flows in the left-hand WLAN are allocated the same air-time for transmitting their payloads, see Figure \ref{fig:ex3}(b).  Observe that the flows in the right-hand WLAN achieve slightly higher throughput and air-time than those in the the left-hand WLAN due to the difference in frame overheads at different PHY rates.


\section{Assumptions}\label{sec:assumptions}

In this section we review the assumptions used in our analysis, and in particular try to identify those assumptions that can be readily relaxed and those that cannot.   Assumption \ref{assump-1} (non-collision losses negligible) can be removed, but see the detailed discussion below.  Assumption \ref{assump0} (homogeneous frame transmission duration) can be readily relaxed to the requirement that stations have the same \emph{mean} frame duration.   Removing this assumption altogether should be possible but requires modifying the denominator (\ref{eq:Xdef}) of the throughput formula to take account of the fact that the duration of a collision now depends on the specific set of stations involved in a collision and so on the attempt rates $x_i$.   Assumption \ref{assump0a} (throughput model) is the fundamental assumption used in our analysis.   This assumption might be weakened in various ways, but is not straightforward to remove.   As discussed in Section \ref{sec:timebased}, it is trivial to remove Assumption \ref{assump1} (homogeneous PHY rates) and so accommodate multi-rate operation and time-based fairness.  Assumptions  \ref{assump1a} and \ref{assump2} can be removed, but similarly to Assumption \ref{assump-1} this is at the cost of a considerable increase in the practical difficulty of realising a max-min allocation.  See the following sections for a detailed discussion, but we note here that perhaps the most notable casualty is that by relaxing these assumptions we lose the equivalence between bottlenecked flows and saturated flows.  This means that standard flow congestion control algorithms (which work by developing a queue backlog) can no longer be relied upon to guarantee flows are bottlenecked.   As already commented upon already, Assumption  \ref{assump2b} can be replaced by a variety of alternative constraints provided we retain log-convexity of the network rate region.   

Lastly, we note that while we have assumed that stations have sufficient arriving traffic to be able to make full use of the max-min fair throughput allocation, our analysis carries over essentially unchanged to situations where the rate of traffic arrivals at stations is itself constrained.   The upper bound on throughput created by the finite traffic load introduces an additional convex constraint, and this constraint becomes the bottleneck when it is less than the max-min fair allocation in the absence of the finite-load constraint.

\subsection{Relaxing Assumption  \ref{assump-1}: non-collision losses negligible}
In this section we consider in more detail what is involved in relaxing Assumption \ref{assump-1}.    The main non-collision sources of loss are channel noise losses, packet discards after too many retries and queue overflow losses.   We begin by noting that excessive channel noise losses can be avoided by appropriate choice of modulation/coding rate, discard losses by use of an appropriate retry limit (the standard value of 11 retries requires a combined channel-noise/collision loss rate exceeding 65\% for the discard probability to exceed 1\%) and queue overflow losses by provisioning links with sufficient buffering.   That is, Assumption \ref{assump-1} can often be satisfied by appropriate network design.   
%
%
When such losses cannot be neglected, more effort is required.  Assume use of a block ACK so that TXOP burst transmissions do not terminate early on detecting a corrupted packet (as they would with per packet ACKing).  This ensures that the duration of TXOP burst transmissions is independent of the specific packet loss pattern experienced by each burst -- the analysis could be extended to include such dependence, but at the cost of a considerable increase in complexity.   TXOP transmissions may consist of multiple blocks destined to different receivers which undergo losses dependent on the receiver.   Under such a model we can use the formulation from \cite{SDL_LossyFair2009}.   Let $s(p)$ now denote the goodput of flow $p$, \emph{i.e.}, the rate received correctly at the destination.   Let $A_{i,p}s(p)$  denote the rate at which station $i$ has to send packets from flow $p$ in order to ensure 
$s(p)$ goodput is received at the destination after undergoing losses at intermediate hops along the route $r(p)$ to the destination.  The scaling term $A_{i,p}\ge 1$ is equal to $1$ if and only if there are no losses along the route from station $i$ to the destination of flow $p$.   Log-convexity of the goodput rate region still holds  and in equations \eqref{eq:step3a} and \eqref{eq:step3c} we now need to replace $s(p)$ with $A_{i,p} s(p)$ to obtain a revised water-filling algorithm that includes the effect of noise losses.  

\begin{figure}[bth]
\centering
\includegraphics[width=0.8\columnwidth]{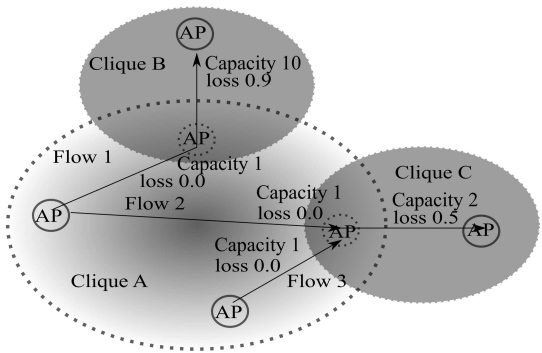}
\caption{Example network with losses to illustrate nature of max-min fair solution.}\label{fig:lossyexample}
\end{figure}

To maintain equal throughput for flows bottlenecked at the same WLAN the station attempt rates $x_k$ have to be adjusted taking into account the term $A_{i,p}$.  When $A_{i,p}$ is not the same for all stations then with per flow burst constraints those bottlenecked flows with smaller values of $A_{i,p}$ will be unsaturated \emph{i.e.} we will lose the equivalence between bottlenecked and saturated flows. We illustrate this with an example. Consider the network in Figure~\ref{fig:lossyexample} where the capacities and the loss rates on the links are chosen such that all of the flows are bottlenecked in WLAN A.  With the restriction that every flow has a maximum burst-size of $1$ (Assumption  \ref{assump1a}), it is easy to see that at the max-min fair solution flow 2 is bottlenecked in clique A but is unsaturated. This is despite the fact that all three flows get the same goodput.    As noted above, fortunately such difficulties can be avoided by the simple expedient of selecting a modulation/PHY rate and retry limit such that losses can be neglected.

\subsection{Relaxing Assumption  \ref{assump1a}: per flow burst-size constraint}
We now consider in more detail removing Assumption \ref{assump1a}. This removes constraint (\ref{eq:step3a}) from the water-filling algorithm and the relaxed optimisation in the proof of Theorem \ref{lem:txop} becomes
\begin{align*}
&\max_{x_k,\bar{x},N_k, n_k} \frac{\bar{x}(c)}{X(c)}\frac{L}{T_c(c)}\\
s.t.\ 
&\bar{x}(c)\le n_kx_k(c)\ \forall k\in \V_\B(c) \\
&\sum_{q\notin\B(c):i(q,c)=k}s(q)\nonumber \\ &\le\frac{N_k x_{k}(c)-|\B(c)\cap\P_k(c)|\bar{x}(c)}{X(c)}\frac{L}{T_c(c)}\ \forall k \\
&1\le N_k(c) \le \bar{N}_k\ \forall\ k\\
&x_k(c)\le y(c)\ \forall k \\
&\prod_{k}(1+x_k)\le \bar{p}(c),\ x_k(c)\ge 0,\ \bar{x}\ge 0\ \forall k 
\end{align*}
where $n_k$ is the burst size used by bottlenecked flows at station $k$ (which must be the same for all bottlenecked flows carried by station $k$ since these flows have the same throughput $\left(n_kx_k/X(c)\right) \left(L/T_c\right)$).   Using similar arguments as those in the proof of Theorem \ref{lem:txop}, the first three constraints will be tight at the optimum.  That is, the burst size $n_k$ will be such that $N_k=\bar{N}_k$ (\emph{i.e.} the maximum admissible value) and the station attempt rate is correspondingly adjusted to maintain $\bar{x}=n_k x_k$.   In general, the burst size $n_k$ and attempt rate $x_k$ will therefore now be different for every station carrying bottlenecked flows (depending on both the number of bottlenecked flows carried by a station and the load imposed by non-bottlenecked flows).    The WLAN attempt rate parameter $y(c)\ge \max_k \bar{x}/n_k$.  Due to the maximisation over $k$ needed here, we may have $x_k<y(c)$ for some stations carrying bottlenecked flows \emph{i.e.} there can exist bottlenecked flows which are unsaturated for all admissible values of $y(c)$ and we lose the equivalence between bottlenecked and saturated flows.  Moreover, it seems clear that stations will generally need to communicate in order to agree the value of $\bar{x}$ and enforce constraint $\bar{x}=n_k x_k$ (equality of bottleneck flow throughputs).  In particular, the selection of $\bar{x}$ is no longer amenable to the decentralised $P_{idle}$ optimisation approach used previously.

\subsection{Relaxing Assumption \ref{assump2}: homogeneous station attempt rate parameters }
Removing Assumption \ref{assump2} removes constraint (\ref{eq:y}) from the relaxed optimisation in the proof of Theorem \ref{lem:txop}.   For stations carrying bottlenecked flows this change has little effect -- all such stations must still use the same attempt rate $\bar{x}$.  In contrast, for stations which carry no bottlenecked flows the attempt rate design parameter can now be selected equal to $x_k$ in which case some of the non-bottlenecked flows will be saturated.   That is, once again we lose the equivalence between bottlenecked and saturated flows.


\section{The hidden terminal issue}
Perhaps the most significant omission from our analysis is hidden terminals.     The basic difficulty here is that we currently lack simple, accurate, generally applicable throughput models when hidden terminals are present, and so we lack the basic tool needed for any max-min fairness analysis.   The modelling difficulty arises from the fact that  hidden terminals can start transmitting even when a transmission by another station has already been in progress for some time.   The class of slotted-time models  pioneered by Bianchi for 802.11 is therefore no longer valid, since these require all transmissions to occur on well-defined MAC slot boundaries, and indeed this  suggests that a fundamental change in modelling paradigm is required.  The development of throughput models in the presence of hidden terminals continues to be the subject of an active research effort, and so in this paper we consider it prudent to leave consideration of utility fairness with hidden terminals to future work. 

It is perhaps also worth noting here that the prevalence of severe hidden terminals in real network deployments presently remains unclear.  While it is relatively easy to construct hidden terminal configurations in the lab that exhibit gross unfairness, it may well be that such configurations are uncommon in practical deployments.  For example, recent measurement studies  report that severe hidden terminal effects typically affect only a relatively small subset of stations in the WLAN deployments considered, \emph{e.g.} see \cite{cheng_Sigcomm_2006,nicu07}.   In mesh network deployments it additionally seems likely that network designers will pro-actively seek to avoid (or at least minimise) creating hidden terminals thereby further reducing their impact.  In addition to appropriate placement of mesh points, hidden terminals can be avoided/mitigated by judicious radio channel assignment and power control (\emph{e.g.} see \cite{colouring} and references therein).  Looking to the future, the latter solutions are facilitated by the trend in next generation networks towards multi-radio architectures and the use of the 5GHz band for mesh backhaul (with its greater number of orthogonal channels compared to the 2.4GHz band).   

Setting the hidden terminal issue to one side for the moment therefore, we stress that the class of mesh networks considered here is a substantial step beyond Aloha, previously the state of the art in wireless utility-fair analysis.   In contrast to Aloha, this class is indeed sufficiently powerful and general to encompass at least some real 802.11 mesh network implementations.  As support for this we comment that we have already implemented one of the max-min fair approaches derived here in an experimental 802.11 testbed using standard hardware and we will report our experimental measurements in due course.


\section{Conclusions}
In this paper we characterise, for the first-time, max-min fair rate allocations for a large class of 802.11 mesh networks.    To our knowledge, this is also the first work to extend max-min fair mesh network analysis beyond Aloha networks.   The class of 802.11 mesh networks considered is large enough to cover realistic network architectures and, by exploiting the features of the 802.11e/n MAC (in particular TXOP packet bursting), we are able to use this characterisation to establish a simple class of network configurations for achieving max-min throughput fairness.   We demonstrate the efficacy of this approach using detailed packet-level simulations and establish that the approach can be readily extended to encompass time-based fairness in multi-rate 802.11 mesh networks.

\section{Acknowledgements}
The authors would like to thank colleague Ken Duffy for his numerous insightful comments and helpful discussions relating to this paper.  

\appendices
\section{Appendix -- Proof of Theorem \ref{lem:txop}}\label{sec:app1}
We proceed by analysing the optimisation (\ref{eq:stepab})-(\ref{eq:step3b}) at step \ref{mainstep} of the water-filling algorithm.    Let $c$ denote a WLAN which becomes a bottleneck at iteration $n$ of the algorithm.    When considering bottlenecked flows at WLAN $c$ we can ignore the constraints at other WLANs  since these constraints must be either loose (or else that WLAN would be the flow bottleneck) or equivalent to the constraints at WLAN $c$ (in the case of a flow having multiple bottlenecks).   Flows which are not bottlenecked at WLAN $c$ must be bottlenecked at other WLANs and the constraints at these WLANs determine the throughput of these flows.
%
%
Let $\B(c)$ denote the set of flows bottlenecked at  WLAN  $c$ and $\V_\B(c)=\{j\in\V: p\in\B(c), p\in \P_j(c)\}$ denote the set of stations carrying one or more bottlenecked flows.   For bottlenecked flows we have that
$$
s(p) = s(q)=T^n=\frac{\bar{x}(c)}{X(c)}\frac{L}{T_c(c)}\ \forall p,q\in \B(c)
$$
for some $\bar{x} \le x_i$ $\forall i \in \V_\B(c)$.  This bottleneck flow throughput is strictly greater than the throughputs of non-bottlenecked flows traversing the WLAN.  By Assumptions \ref{assump0}-\ref{assump1a} all flows $q$ satisfy $s(q) \le \frac{{x}_k(c)}{X(c)}\frac{L}{T_c(c)}$.   Let us relax, for the moment, equality in (\ref{eq:one}) and replace it by the RHS upper bounding the LHS.   By Assumption \ref{assump2} all stations use the same attempt probability design parameter $y(c)$ and $x_k(c)\le y(c)$ for every station $k$.
Combining these observations, leads us to consider the following relaxed optimisation problem,
\begin{align}
&\max_{x_k,\bar{x},N_k} \frac{\bar{x}(c)}{X(c)}\frac{L}{T_c(c)}\label{eq:costfn}\\
s.t.\ 
&\bar{x}(c)\le x_k(c)\ \forall k\in \V_\B(c) \label{eq:bot}\\
&s(q) \le \frac{{x}_k(c)}{X(c)}\frac{L}{T_c(c)}\ \forall q\in\{p\notin \B(c) :i(p,c)=k\}, \forall\ k  \label{eq:ineq2a}\\
&\sum_{q\notin\B(c):i(q,c)=k}s(q)\nonumber \\ &\le\frac{N_k x_{k}(c)-|\B(c)\cap\P_k(c)|\bar{x}(c)}{X(c)}\frac{L}{T_c(c)}\ \forall k \label{eq:ineq3}\\
&1\le N_k(c) \le |\P_k(c)|\ \forall\ k\\
&x_k(c)\le y(c)\ \forall k \label{eq:y}\\
&\prod_{k}(1+x_k)\le \bar{p}(c),\ x_k(c)\ge 0,\ \bar{x}\ge 0\ \forall k \label{eq:pidle0}
\end{align}
where $i(p,c)$ denotes the access point relaying flow $p$ on channel $c$.  It can be verified that this relaxed optimisation can be transformed into a convex problem and so has a unique solution\footnote{Change variables to $\log x_k$, $\log N_k$ and $\log \bar{x}$.   $X$ is a posynomial and so when expressed in terms of these transformed variables $\log X$ is the log sum of exponentials and convex.}.   

Consider the following constraints on station $k\in\V_\B(c)$ carrying at least one bottlenecked flow,
\begin{align}
\bar{x}(c)&\le x_k(c) \label{eq:ineq1}  \\
s(q) &\le \frac{ x_k(c)}{X(c)}\frac{L}{T_c(c)}\ \forall q\in\{p\notin\B(c):i(p,c)=k\}  \label{eq:ineq2} \\
x_k(c)&\le y(c) \label{eq:ineq2b} 
\end{align}
The last constraint is satisfied provided $y(c)\ge \bar{x}$ -- we return to the choice of $y(c)$ shortly.  It can be verified (\emph{e.g.} by inspecting derivatives with respect to $x_k$) that $x_k$ and $x_k/X$ are strictly increasing in $x_k$, while $1/X$ is strictly decreasing in $x_k$.    Hence, if inequalities (\ref{eq:ineq1}) and (\ref{eq:ineq2})  are both loose then decreasing $x_k(c)$ decreases the RHS while improving the cost function and making the other inequalities looser, which leads to a contradiction.  Hence we may must have equality in either/both (\ref{eq:ineq1}) and (\ref{eq:ineq2}) (for at least one $q$).  Recalling that for non-bottlenecked flows $s(q)< \frac{\bar{x}}{X(c)}\frac{L}{T_c(c)}\le \frac{x_k}{X(c)}\frac{L}{T_c(c)}$, it can be seen that constraint (\ref{eq:ineq1}) will always become tight before constraint (\ref{eq:ineq2}).   Hence, we must have equality in (\ref{eq:ineq1}).  That is, $x_k=\bar{x}$ for all stations carrying a bottlenecked flow and  for any bottlenecked flow the burst-size used is exactly one frame per successful transmission by the station.  For non-bottlenecked flows the average burst size per successful transmission by the station must be strictly less than one frame, which implies that these flows are unsaturated.
 
Turning now to station $k\notin\V_\B(c)$ that carries no bottlenecked flows, constraint (\ref{eq:ineq1}) no longer applies but (\ref{eq:ineq2b}) and (\ref{eq:ineq2}) are still in force. Since all flows on the station are, by assumption, non-bottlenecked they have throughput strictly less than $\frac{\bar{x}}{X(c)}\frac{L}{T_c(c)}$.  Hence, if we have equality in (\ref{eq:ineq2}) for one or more flows then $x_k(c)<\bar{x}$.  But from (\ref{eq:ineq2b}) and the discussion in the foregoing paragraph $y(c)\ge\bar{x}$ and so $x_k(c)<\bar{x}\le y$.    Since $x_k<y$ the station is unsaturated and therefore also every flow is unsaturated.  If we have inequality in (\ref{eq:ineq2}) for all flows then the average flow burst size must be strictly less than one frame per successful transmission by the station which implies that, once again, every flow is unsaturated.

To gain insight into the burst size $N_k$, we need to consider constraint (\ref{eq:ineq3}).   Since $N_kx_k/X$ is increasing in $N_k$ and $1/X$ is decreasing, using a similar contradiction argument as previously we must have equality in (\ref{eq:ineq3}) for all stations.


Consider now the value of $y(c)$.  It can be seen that $x_k(c)$ is invariant in $y(c)\ge\bar{x}$.  Hence any $y(c)\ge\bar{x}$ is an admissible solution and yields the same allocation of $x_k$'s and $N_k$'s.  Since we have equality in  (\ref{eq:ineq3}), these solutions to the relaxed optimisation are also feasible for the true/unrelaxed constraints.   Observe, however, that when $y>\bar{x}$ no flow is saturated (for stations $k\in\V_\B(c)$, $x_k=r_k(y)=\bar{x}<y$ and so the stations are unsaturated and thus every flow must be unsaturated, for stations $k\notin\V_\B(c)$ we already have that every flow is unsaturated).   When $y=\bar{x}$ we have that all bottlenecked flows are saturated and all non-bottlenecked flows are unsaturated (for stations $k\in\V_\B(c)$, $x_k=r_k(y)=\bar{x}=y$ and so the station is saturated plus bottlenecked flows send one packet at every successful transmission by a station and so are also saturated since a flow cannot know in advance which transmissions will be successful, for all stations we already have that non-bottlenecked flows are unsaturated).   Observe  also that while we have some freedom in the choice of $y$, since the max-min allocation for the original problem is unique the values of the $x_k$'s and $N_k$'s (which are invariant in $y\ge\bar{x}$) are unique.     

{}

\begin{biography}{Doug Leith} graduated from the University of Glasgow in 1986 and was awarded his PhD, also from the University of Glasgow, in 1989. In 2001, Prof. Leith moved to the National University of Ireland, Maynooth to assume the position of SFI Principal Investigator and to establish the Hamilton Institute (www.hamilton.ie) of which he is Director.  His current research interests  include the analysis and design of network congestion control and  resource allocation in wireless networks.
\end{biography}

\begin{biography}{Qizhi Cao} 
\end{biography}

\begin{biography}{Vijay Subramanian} received the B.Tech. degree from IIT Madras,in 1993, the M.S. degree from the Indian Institute of Science, Bangalore, in 1995, and the Ph.D. degree from the University of Illinois at Urbana-Champaign, Urbana, in 1999. From 1999 to 2006, he was with the Networks Business, Motorola, Arlington Heights, IL where he worked on developing wireless scheduling algorithms deployed in many of Motorola's wireless data products. Since May 2006 he is a Research Fellow at the Hamilton Institute, NUIM, Ireland. His research interests include information theory, communication networks, queueing theory, mathematical immunology and applied probability.
\end{biography}

\end{document}